\documentclass[apj]{emulateapj}
\usepackage{mathptmx}

\def\gtorder{\mathrel{\raise.3ex\hbox{$>$}\mkern-14mu
             \lower0.6ex\hbox{$\sim$}}}
\def\ltorder{\mathrel{\raise.3ex\hbox{$<$}\mkern-14mu
             \lower0.6ex\hbox{$\sim$}}}

\slugcomment{Accepted tp PASP}

\shorttitle{PTF calibrated photometry}

\shortauthors{Ofek et al.}

\begin{document}

\title{The Palomar Transient Factory photometric calibration}
\author{
E.~O.~Ofek\altaffilmark{1}$^{,}$\altaffilmark{2}$^{,}$\altaffilmark{13},
R.~Laher\altaffilmark{3},
N.~Law\altaffilmark{4},
J.~Surace\altaffilmark{3},
D.~Levitan\altaffilmark{1},
B.~Sesar\altaffilmark{1},
A.~Horesh\altaffilmark{1},
D.~Poznanski\altaffilmark{5}$^{,}$\altaffilmark{6}$^{,}$\altaffilmark{2},
J.~C.~van~Eyken\altaffilmark{7},
S.~R.~Kulkarni\altaffilmark{1},
P.~Nugent\altaffilmark{5},
J.~Zolkower\altaffilmark{8},
R.~Walters\altaffilmark{8},
M.~Sullivan\altaffilmark{9},
M.~Ag\"{u}eros\altaffilmark{10},
L.~Bildsten\altaffilmark{11}$^{,}$\altaffilmark{12}
J.~Bloom\altaffilmark{6},
S.~B.~Cenko\altaffilmark{6},
A.~Gal-Yam\altaffilmark{13},
C.~Grillmair\altaffilmark{3},
G.~Helou\altaffilmark{3},
M.~M.~Kasliwal\altaffilmark{1},
R.~Quimby\altaffilmark{1}
}
\altaffiltext{1}{Division of Physics, Mathematics and Astronomy, California Institute of Technology, Pasadena, CA 91125, USA}
\altaffiltext{2}{Einstein Fellow.}
\altaffiltext{3}{Spitzer Science Center, MS 314-6, California
  Institute of Technology, Jet Propulsion Laboratory, Pasadena, CA
  91125.}
\altaffiltext{4}{Dunlap Institute for Astronomy and Astrophysics,
  University of Toronto, 50 St. George Street, Toronto, Ontario M5S
  3H4, Canada.}
\altaffiltext{5}{Lawrence Berkeley National Laboratory, 1 Cyclotron
  Road, Berkeley, CA 94720.}
\altaffiltext{6}{Department of Astronomy, University of California,
  Berkeley, Berkeley, CA 94720-3411.}
\altaffiltext{7}{NASA Exoplanet Science Institute, California Institute of Technology, Pasadena, CA 91125, USA}
\altaffiltext{8}{Caltech Optical Observatories, California Institute
  of Technology, Pasadena, CA 91125.}
\altaffiltext{9}{Department of Physics, University of Oxford, Denys
  Wilkinson Building, Keble Road, Oxford OX1 3RH, UK.}
\altaffiltext{10}{Columbia University, Department of Astronomy, 550 West 120th street, New York, NY 10027}
\altaffiltext{11}{Department of Physics, Broida Hall, University of
  California, Santa Barbara, CA 93106.}
\altaffiltext{12}{Kavli Institute for Theoretical Physics, Kohn Hall,
  University of California, Santa Barbara, CA 93106.}
\altaffiltext{13}{Benoziyo Center for Astrophysics, Weizmann Institute
  of Science, 76100 Rehovot, Israel.}

\begin{abstract}

The Palomar Transient Factory (PTF) provides multiple epoch imaging
for a large fraction of the celestial sphere.
Here we describe the photometric calibration
of the PTF data products
that relates the PTF magnitudes
to other magnitude systems.
The calibration process utilizes Sloan Digital Sky Survey (SDSS)
$r\sim16$\,mag point source objects as photometric standards.
During photometric conditions, this allows us to solve for
the extinction coefficients and color terms, and to estimate the camera
illumination correction.
This also enables the calibration of fields
that are outside the SDSS footprint.
We test the precision and repeatability of the PTF
photometric calibration.
Given that PTF is observing in a single filter
each night, we define a PTF calibrated magnitude system 
for $R$-band and $g$-band.
We show that, in this system, $\approx 59\%$ ($47\%$)
of the photometrically calibrated PTF $R$-band ($g$-band) data
achieve a photometric precision
of 0.02--0.04\,mag,
and have color terms and extinction coefficients
that are close to their average values.
Given the objects' color, the PTF magnitude system
can be converted to other systems.
Moreover,
a night-by-night
comparison of the calibrated magnitudes of individual stars
observed on multiple nights
shows that they are consistent to a level of
$\approx0.02$\,mag.
Most of the data that were taken under non-photometric
conditions can be calibrated relative to other epochs
of the same sky footprint obtained during photometric
conditions.
We provide a concise guide describing the use of the PTF
photometric calibration data products,
as well as
the transformations between the PTF magnitude system
and the SDSS and Johnson-Cousins systems.

\end{abstract}


\section{Introduction}
\label{Introduction}

The Palomar Transient
Factory\footnote{http://www.astro.caltech.edu/ptf/}
(PTF; Law et al. 2009; Rau et al. 2009)
is a synoptic survey designed to explore the transient sky.
The project utilizes the $48''$ Samuel Oschin Schmidt Telescope
on Mount Palomar.
The telescope has a digital camera equipped with 11 active CCDs, each 2K$\times$4K pixels
(Rahmer et al. 2008).
Each PTF image covers 7.26\,deg$^{2}$
with a pixel scale of $1.01''$\,pix$^{-1}$.
From the beginning of the PTF survey in March 2009 until
January 2011, most of the images were taken with
the Mould $R$ filter.
Starting January 2011, we performed the
PTF main survey in $g$-band
during dark time and in $R$-band during bright time.
In addition, a few nights around times of full Moon
are used for taking images with narrow-band  H$\alpha$
filters.
A PTF system overview and review of the first year's performance are
given in Law et al. (2010).

Accurate photometric calibration is a non-trivial task since
one must know both the atmospheric transmission
(e.g., Padmanabhan et al. 2008; Burke et al. 2010)
and the optical/detector system
response as a function of wavelength, time,
and sky position.
Moreover, observations of flux standards are required, and the
true spectral energy distribution of these standards
needs to be known.
In practice 
simplifying assumptions are made
in order to achieve solutions which are good to
the few percent level.
For example, in many cases, it is customary to use the calibrated magnitudes
of the standard stars instead of their
spectral energy distribution,
and to assume that the atmospheric transmission is a
smooth function of airmass.
We note that 
relative-photometry approaches
(e.g., Gilliland \& Brown 1988;
Gilliland et al. 1991; Honeycutt 1992;
Everett \& Howell 2001;
Ofek et al. 2011a)
can achieve better
accuracy from the ground, but are limited by
scintillation noise (e.g., Young 1967; Gilliland \& Brown 1988),
Poisson noise and flat-fielding errors.

There are several approaches to photometric calibration.
For example, the Sloan Digital Sky Survey (SDSS; York et al. 2000)
images are calibrated using an auxiliary 20-inch telescope,
which determines the photometric condition on a nightly basis (Hogg et al. 2001).
This telescope measures the extinction and photometric
zero point using a network of standard stars (Smith et al. 2002),
which, in turn, are tied to the standard star BD$+17$:$4708$.
Furthermore, the 
photometric uniformity in the SDSS
is achieved by
the algorithm described in Padmanabhan et al. (2008).
The SDSS photometry is uniform to better than 2\% in
all bands (Adelman-McCarthy et al. 2008).


Ofek (2008; see also Pickles \& Depagne 2010)
has suggested using Tycho-2 (H{\o}g et al. 2000)
stars to photometrically calibrate astronomical images.
The SDSS $griz$ magnitudes of such stars are based on
synthetic magnitudes of stellar spectral templates fitted
with the Hipparchos $B_{T}$, $V_{T}$
magnitudes and the 2MASS (Skrutskie et al. 2006)
$JHK$ magnitudes.
However, this strategy requires images
containing unsaturated Tycho calibration stars
which lie in the $\approx9$--$12$ magnitude range.
The saturation limit for the PTF $g$- and $R$-band survey
is typically\footnote{The PTF camera electronics were modified on 2009 Oct 22
to increase the dynamic range. Before this date, the saturation limit
was around 15th magnitude.}
around 14 magnitude.
%

Herein, we 
describe the method we have developed for the
photometric calibration of PTF data
taken in the $g$ and $R$ bands.
The photometric calibration of the PTF H$\alpha$ survey will be described elsewhere.
We note that the relative photometric calibration of PTF
currently\footnote{For relative photometry, we use a scheme
similar to that proposed by Honeycutt (1992) with some modifications outlined in Ofek et al. (2011a).}
achieves precision as good as 3\,mmag, in given fields, at magnitude 15
(e.g., van Eyken et al. 2011; Ag\"{u}eros et al. 2011; Law et al. 2011; Levitan et al. 2011;
Polishook et al. 2011).
The PTF relative photometry pipeline
will be described in Levitan et al. (in prep.).

Our method of photometric calibration for
PTF data
is similar to the ``classical'' method of observing
a small number of calibration stars (e.g., Landolt 1992)
through various airmasses and assuming photometric conditions --
i.e., the atmosphere transmission properties are constant in time
and are a continuous function of airmass.
The main difference is that we are using SDSS stars
as standard stars
and we 
typically observe $\sim10^{5}$ SDSS stars with high S/N
in each CCD per night.
Another important difference is that PTF observations
are done in a single filter each night.
Therefore, in order to relate the PTF calibrated magnitudes
to one of the common absolute systems one need to
apply color terms.
The paper is organized as follows.
In \S\ref{Process}, we describe the PTF photometric-calibration method.
In \S\ref{Illum}, we discuss the illumination correction,
while \S\ref{fog} describes a related problem that plagues early PTF data.
In \S\ref{Data} we describe the data products and 
what calibration information is stored with the image-product data.
The performance of the PTF magnitude calibration is given in \S\ref{Perf},
and the 
derived photometric parameter statistics are discussed in \S\ref{Pars}.
We provide the color transformation between the PTF magnitude system
and other systems in \S\ref{Color}.
Finally, we conclude the paper in \S\ref{Sum}.
Unless specified otherwise, the statistics given here are
based on all PTF data obtained from March 2009 to July 2011.

\section{Photometric-calibration method}
\label{Process}

PTF has two main data-reduction pipelines.
The first is for
real-time ($\approx30$\,min; e.g., Gal-Yam et al. 2011)
image subtraction 
and transient detection, hosted by the Lawrence Berkeley National Laboratory
(Nugent et al., in prep.).
This paper will make no further mention of this pipeline, as
only the second pipeline is relevant here, as discussed below.

The photometric calibration described in this paper
is implemented in the second pipeline.
This pipeline, hosted by the
Infrared Processing and Analysis Center (IPAC),
performs final image reduction
and extracts the source catalogs.
The processing includes splitting
the multi-extension FITS images, de-biasing, flat fielding,
astrometric calibration,
generation of mask images, 
source extraction
and photometric calibration (the subject of this paper).
This pipeline is described in 
Grillmair et al. (2010) and Laher et al. (in prep.).

Our photometric-calibration process
runs on PTF data separately for each night, filter and CCD.
It first attempts
to match the sources extracted from
the PTF images taken during a given night
with SDSS-DR7 PhotoPrimary\footnote{PhotoPrimary is an SDSS table. See definitions in the SDSS Schema Browser: http://cas.sdss.org/dr7/en/help/browser/browser.asp}
point sources (i.e., SDSS type$=6$).
In order to assure good photometric quality, only SDSS
stellar objects 
with photometric errors smaller
than 0.05\,mag in $r$- and $i$-bands
and which are fainter than 15 mag (to avoid saturated stars)
in the $g$-, $r$- and $i$-bands are used.
The photometric solutions
are calculated only if more than 30 science images
were taken during the night,
and only if we were able to select more than 1000
SDSS stars in all the images taken with a given CCD and filter
during the entire night.

We use the SDSS-matched stars as a set of
standard stars and
solve for the photometric zero points,
airmass terms and color terms in a given night.
The fitting process is done separately for each
one of the 11 active CCDs and the $g$- and $R$-band filters.
This is required since
the CCDs are not identical and some of them have
a different spectral response (see Law et al. 2009 for details).
For observations taken using the
$R$-band\footnote{The PTF Mould $R$ filter is similar in shape
to the SDSS $r$-band filter, but shifted 27\,\AA~redward.}
filter we fit the following model:
\begin{eqnarray}
r_{{\rm SDSS}}-R_{{\rm PTF}}^{{\rm inst}}  & =  ZP_{R} + \alpha_{{\rm c},R} (r_{{\rm SDSS}}-i_{{\rm SDSS}}) \cr 
                 &    + \alpha_{{\rm a},R}AM + \alpha_{{\rm ac},R}AM(r_{{\rm SDSS}}-i_{{\rm SDSS}}) \cr
                 &    + \alpha_{{\rm t},R}(t-t_{{\rm m}}) + \alpha_{{\rm t2},R}(t-t_{{\rm m}})^{2} \cr 
                 &    - 2.5\log_{10}(\delta{t}),
\label{ZPeq_R}
\end{eqnarray}
while for $g$-band observations we fit:
\begin{eqnarray}
g_{{\rm SDSS}}-g_{{\rm PTF}}^{{\rm inst}}  & =  ZP_{g} + \alpha_{{\rm c},g} (g_{{\rm SDSS}}-r_{{\rm SDSS}}) \cr 
                 &    + \alpha_{{\rm a},g}AM + \alpha_{{\rm ac},g}AM(g_{{\rm SDSS}}-r_{{\rm SDSS}}) \cr
                 &    + \alpha_{{\rm t},g}(t-t_{{\rm m}}) + \alpha_{{\rm t2},g}(t-t_{{\rm m}})^{2} \cr 
                 &    - 2.5\log_{10}(\delta{t}).
\label{ZPeq_g}
\end{eqnarray}
%
%
Here, $f_{{\rm PTF}}^{{\rm inst}}$ 
is the PTF instrumental magnitudes in band $f$ (either $R$ or $g$),
$f_{SDSS}$ is the SDSS magnitude in band $f$ (either $g$, $r$ or $i$),
$ZP_{f}$ is the photometric zero point for filter $f$,
$\alpha_{{\rm c},f}$ is the color term for filter $f$,
$\alpha_{{\rm a},f}$ is the extinction coefficient
(airmass term) for filter $f$, 
$\alpha_{{\rm ac},f}$ is the airmass-color term for filter $f$,
$AM$ is the airmass,
$\alpha_{{\rm t},f}$ and
$\alpha_{{\rm t2},f}$ are the polynomial coefficients for
the change in the zero point of filter $f$ as a function of time, $t$ in days,
during the night,
where $t_{{\rm m}}$ in days is the middle of the night,
and $\delta{t}$ is the exposure time in seconds.
%

The instrumental magnitudes
used in the photometric calibration process
are based on the SExtractor (Bertin \& Arnouts 1996)
MAG\_AUTO magnitude\footnote{Defined with ${\rm kron\_fact}=1.5$ and ${\rm min\_radius}=2.5$.} (see also \S\ref{Acc}), with
an internal SExtractor zero point of 0.
The above set of equations are solved using linear least-squares fitting.
The errors in $r_{{\rm SDSS}}-R_{{\rm PTF}}^{{\rm inst}}$
are taken as $(\Delta{r_{{\rm SDSS}}}^{2}+\Delta{R_{{\rm PTF}}^{{\rm inst}}}^{2}+0.015^{2})^{1/2}$,
where $\Delta{r_{{\rm SDSS}}}$ is the SDSS magnitude error, $\Delta{R_{{\rm PTF}}^{{\rm inst}}}$
is the PTF magnitude errors and $0.015$ is the assumed
internal accuracy of the SDSS
photometric calibration
(Adelman-McCarthy et al. 2008).
The fit is performed iteratively,
up to 3 times, with sigma clipping of 3$\sigma$.
The sigma clipping ensures removal of stars with
bad photometry (e.g., influenced by cosmic rays; saturated pixels).

Next, in order to be able to correct for zero-point variations
across a given CCD (see \S\ref{Illum}),
the residuals from the best fit of Equations~\ref{ZPeq_R} or \ref{ZPeq_g} are binned in
cells of $256\times256$\,pix$^{2}$ along the X and Y dimensions
of each CCD. In each cell, we take the mean of the
residuals and subtract from it
the mean of the residuals in the center-of-image cell\footnote{The position of the center of this central cell is $x=1025$, $y=2049$.},
in order to render the residuals relative to the residual at the image center.
The resulting coarse image of the mean of the residuals is
linearly interpolated
to generate an image at the resolution of PTF images
of pixel-to-pixel zero point variations.
This image product is equivalent to an illumination
correction, and herein also called the zero point variation map (ZPVM; see \S\ref{Illum}).
The uncertainty in the ZPVM
is estimated by calculating
the standard deviation (StD) of the residuals in each
cell\footnote{The StD is not divided by the square root of number of
data points in each cell. Therefore, it represents the scatter rather than the error in the mean.}.

We note that the PTF illumination correction images are usually smooth
on large scales (comparable with the CCD image size).
Therefore, in most cases the ZPVM can be represented
by low-order polynomials.
In order to provide users with a simpler version of
the illumination correction, we also fit
versions of Equations~\ref{ZPeq_R} and \ref{ZPeq_g} that consists of
a low-order-polynomial representation of the ZPVM, e.g., for the $R$-band:
\begin{eqnarray}
r_{{\rm SDSS}}-R_{{\rm PTF}}^{{\rm inst}}  & = ZP_{R} + \alpha_{{\rm c},R} (r_{{\rm SDSS}}-i_{{\rm SDSS}}) \cr
                 &    + \alpha_{{\rm a},R}AM + \alpha_{{\rm ac},R}AM(r_{{\rm SDSS}}-i_{{\rm SDSS}}) \cr
                 &    + \alpha_{{\rm t},R}(t-t_{{\rm m}}) + \alpha_{{\rm t2},R}(t-t_{{\rm m}})^{2} \cr
                 &    - 2.5\log_{10}(\delta{t})\cr
                 &    + \alpha_{{\rm x1},R}(X-\frac{1}{2}X_{{\rm size}})/X_{{\rm size}}               \cr
                 &    + \alpha_{{\rm y1},R}(Y-\frac{1}{2}Y_{{\rm size}})/Y_{{\rm size}}               \cr
                 &    + \alpha_{{\rm y2},R}[(Y-\frac{1}{2}Y_{{\rm size}})/Y_{{\rm size}}]^{2}         \cr
                 &    + \alpha_{{\rm y3},R}[(Y-\frac{1}{2}Y_{{\rm size}})/Y_{{\rm size}}]^{3}         \cr
                 &    + \alpha_{{\rm xy},R}(X-\frac{1}{2}X_{{\rm size}})(Y-\frac{1}{2}Y_{{\rm size}})/(X_{{\rm size}} Y_{{\rm size}}),
\label{ZPeqXY}
\end{eqnarray}
where $X$ and $Y$ are the positions on the CCD,
$X_{{\rm size}}$ is the size of the CCD, in pixels, in the $X$ dimension (2048\,pix)
and $Y_{{\rm size}}$ corresponds to the $Y$ dimension (4096\,pix).
A similar equation is fitted for the $g$-band observations
(see Equation~\ref{ZPeq_g} for analogy).
Here, the illumination correction is represented by polynomials
with coefficients $\alpha_{{\rm x1},f}$,
$\alpha_{{\rm y1},f}$, $\alpha_{{\rm y2},f}$, $\alpha_{{\rm y3},f}$,
and $\alpha_{{\rm xy},f}$.
This representation usually provides a good estimate of the illumination
correction over the image.
We note that for each solution we store a variety of information
regarding the quality of the fit (see \S\ref{Data}).

\section{Illumination correction}
\label{Illum}

\begin{figure}
\centerline{\includegraphics[width=8.5cm]{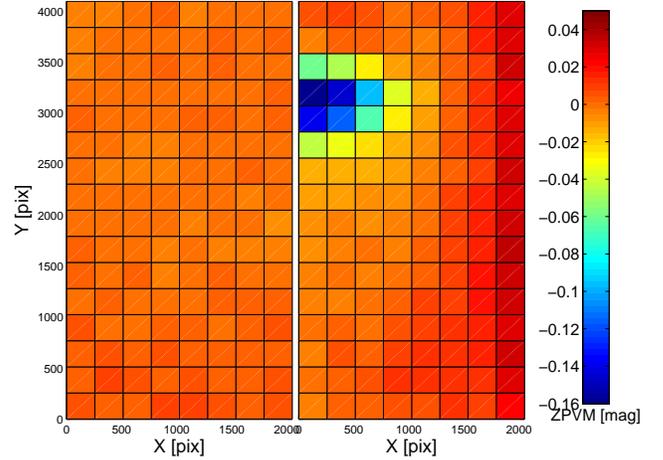}}
\caption{Two examples of ZPVM images. The color coding represents the
variations in photometric zero point
with respect to the center of image. {\it Left panel:}
a good case with peak-to-peak range of 0.016\,mag
in the zero point variations across the image. This ZPVM image was constructed based
on data obtained on 2011 July 13 with CCDID$=8$ and the $R$-band filter.
{\it Right panel:} One of the worst examples with
peak-to-peak range of 0.22\,mag
in the zero point across the image.
This ZPVM image was constructed based
on data obtained on 2010 May 7 with CCDID$=5$ and the $R$-band filter.
In such bad cases the polynomial representation may
deliver poor quality results.
We note that this bad example is not representative of PTF data
and that typical images obtained using this system have good image quality.
\label{fig:ZPVM_example}}
\end{figure}

The ZPVM and its polynomial
representation amount to an illumination correction that
describes the variations in the photometric zero point
spatially across the image (e.g., Regnault et al. 2009).
Naively, such zero-point variations should be removed by
the flat-fielding process.
However, for example, if the fraction of flux in the stars' point spread function (PSF)
wings, relative to the flux integrated to infinity in the PSF,
is variable as a function of position on the CCD
then this may induce variations that are not removed
by the flat-fielding process.
Such CCD position-dependent variations are indeed
detected in PTF images (e.g., \S\ref{fog}).
Figure~\ref{fig:ZPVM_example} presents two examples of illumination
correction images. The left panel shows
a typical case, in which the illumination
corrections have a low range variations.
The right panel shows one of the worst cases we encountered
thus far,
with larger amplitude variations (see \S\ref{fog}).
We note that the median peak-to-peak variations
in the zero point across the image over all photometric nights,
filters and CCDs is about 0.06\,mag.
%



In the first year of the PTF project
following first light, we detected a problem
with the PTF image quality.
This issue is referred to as the fogging problem,
and it manifested itself as a diffuse halo around bright stars
and a prominent large scale non-uniformity in
the illumination corrections.
The fogging problem and our
testing and subsequent apparatus modifications for 
ameliorating its effects are described in \S\ref{fog}.

\section{The fogging problem}
\label{fog}

The PTF camera window,
which also acts as a field flattener, is fairly large
($\approx320$\,cm$^{2}$).
The window surface temperature is $\approx16$\,$^{\circ}$C below the ambient
temperature.  To prevent condensation from forming on the window, a constant
stream of dry air or nitrogen is blown across
the window's outer surface.  In the
original installation of the camera, new nylon tubing was used to convey the
gas to the window.  This tubing was selected for its reported high
durability, flexibility and low outgassing properties.  Once the window
fogging problem was discovered, a test chamber was assembled, which simulated the
cold window environment in which dry nitrogen could be conveyed across a
cold test surface.

The outgassing properties of various tubing materials
were evaluated in the test chamber.
The nylon tubing used in the
original camera installation was found to have a moderate amount of
outgassing which produced volatiles that would condense on the cold test
surface.  Tubing made from Teflon FEP (fluorinated ethylene propylene) was
found to have low outgassing with no detectable volatile condensation on the
cold test surface.
The camera window was cleaned  and all tubing between the dry nitrogen supply and the
camera window was replaced with this material on 2010 September 2.
No window fogging issues have been observed since the tubing
material was replaced.

We note that the camera window was cleaned several
times during PTF operations, and after each cleaning
an immediate reduction
in ZPVM deviations from zero is expected.
The cleaning dates are listed in Table~\ref{Tab:Cleaning}.
\begin{deluxetable}{llll}
\tablecolumns{4}
\tabletypesize{\scriptsize}
\tablewidth{0pt}
\tablecaption{PTF camera window cleaning dates}
\tablehead{
\multicolumn{3}{c}{Cleaning date} &
\colhead{} \\
\colhead{Year} &
\colhead{Month} &
\colhead{Day} &
\colhead{Comments} 
}
\startdata
2009 & 07 & 17 & \\
2009 & 07 & 29 & \\
2009 & 08 & 11 & \\
2009 & 08 & 20 & \\
2009 & 09 & 15 & \\
2009 & 09 & 29 & \\
2009 & 10 & 21 & \\
2009 & 11 & 10 & \\
2010 & 01 & 11 & \\
2010 & 01 & 27 & \\
2010 & 02 & 04 & \\
2010 & 02 & 18 & \\
2010 & 03 & 04 & \\
2010 & 04 & 06 & \\
2010 & 04 & 21 & \\
2010 & 05 & 12 & \\
2010 & 06 & 02 & \\
2010 & 06 & 17 & \\
2010 & 07 & 01 & \\
2010 & 08 & 04 & \\
2010 & 08 & 16 & \\
2010 & 09 & 02 & new tubing
\enddata
\tablecomments{Local dates the camera window was cleaned
(e.g., the last cleaning date is 2010 Sep 02, so all the data
taken on UTC 2010 Sep 03 and afterward is affected by this cleaning).
At these dates, the illumination correction may change markedly.}
\label{Tab:Cleaning}
\end{deluxetable}
Fortunately, the method described in \S\ref{Process}
provides an estimate of the zero-point variations
across the CCD image.
Therefore, the illumination correction
is used to correct our photometry,
and to monitor the PTF image quality.

\section{Calibration Products}
\label{Data}

The parameters required for calibrating
the instrumental magnitudes
are stored in the PTF database, as well
as in the image and catalog FITS file headers (\S\ref{PrHead}).
In addition, a full version of each illumination-correction image 
is stored as ancillary calibration data in a FITS file,
and associated metadata are stored in the PTF database (\S\ref{ZPVM}).
The use of these data products is discussed in \S\ref{Use}.
%

\subsection{FITS headers}
\label{PrHead}

The parameters based on the solution of Equation~\ref{ZPeqXY}
are stored in the image headers.
Moreover,
generated from each image is a
FITS binary table
containing all the sources extracted from that image using
SExtractor.
The FITS binary table catalogs
contain three $\rm{header} + \rm{data}$ units (HDUs) and two FITS extensions.
The first HDU (i.e., the primary HDU) contains a header with
metadata from SExtractor and no data.  The second HDU, or first FITS
extension, is a binary table containing the properties of the sources
extracted from the processed image, with its header
giving information about the table columns.
The source properties in the binary table include aperture-photometry 
instrumental magnitudes and other SExtractor outputs.
The binary table has also been augmented with an additional column 
called ZEROPOINT, which stores the source-dependent zero points 
computed from our photometric-calibration parameters; these zero points 
already include the $2.5 \log( \delta t)$ contribution for normalizing 
the image data by the exposures time, 
$\delta t$, in seconds, and so straightforwardly adding the instrumental
magnitudes to their respective zero points
will result in calibrated magnitudes.
The third HDU, or second FITS
extension, contains an empty image whose header is a copy of the
processed-image header, which includes information about the
photometric calibration as previously mentioned.
The header keywords related to the photometric solutions are summarized
in Table~\ref{Tab:Header}. 
%
\begin{deluxetable*}{llll}
\tablecolumns{4}
\tabletypesize{\scriptsize}
\tablewidth{0pt}
\tablecaption{The photometric calibration FITS header keywords}
\tablehead{
\colhead{Keyword}   &
\colhead{Symbol}  &
\colhead{Units}  &
\colhead{Description} 
}
\startdata
FILTER   & $f$                 &      & Filter name (e.g., $g$, $R$) \\
AEXPTIME &$\delta{t}$          & s    & Actual exposure time\\
APSFILT  &                     &      & SDSS filter used in photometric calibration (e.g., $g$, $r$)\\
APSCOL   &                     &      & SDSS color used in photometric calibration (e.g., $g-r$, $r-i$)\\
APRMS    &                     & mag  & RMS of the residuals in photometric calibration \\
APBSRMS  &                     & mag  & RMS of the residuals in photometric calibration for stars brighter than 16th magnitude in $r$-band.\\
APNSTDI1 &                     &      & Number of calibration stars in first iteration   \\
APNSTDIF &                     &      & Number of calibration stars in final iteration   \\
APCHI2   &                     &      & $\chi^{2}$ of photometric calibration\\
APDOF    &                     &      & Number of degrees of freedom in photometric calibration \\
APMEDJD  &$t_{m}$              & day  & Median JD for the night\\
APPN01   &                     &      & Name of parameter photometric calibration 01 (i.e., 'ZeroPoint')\\
APPAR01  &$ZP_{f}$             & mag\,ADU$^{-1}$\,s$^{-1}$  & Value of parameter photometric calibration 01\\
APPARE01 &$\Delta{ZP_{f}}$     & mag  & Error in parameter photometric calibration 01\\
APPN02   &                     &      & Name of parameter photometric calibration 02 (i.e., 'ColorTerm')\\
APPAR02  &$\alpha_{{\rm c},f}$  & mag\,mag$^{-1}$ & Value of parameter photometric calibration 02\\
APPARE02 &$\Delta{\alpha_{{\rm c},f}}$ & mag\,mag$^{-1}$ & Error in parameter photometric calibration 02\\
APPN03   &                     &      & Name of parameter photometric calibration 03 (i.e., 'AirMassTerm')\\
APPAR03  &$\alpha_{{\rm a},f}$         & mag\,airmass$^{-1}$ & Value of parameter photometric calibration 03\\
APPARE03 &$\Delta{\alpha_{{\rm a},f}}$ & mag\,airmass$^{-1}$ & Error in parameter photometric calibration 03\\
APPN04   &                     &      & Name of parameter photometric calibration 04 (i.e., 'AirMassColorTerm')\\
APPAR04  &$\alpha_{{\rm ac},f}$        & mag\,mag$^{-1}$\,airmass$^{-1}$ & Value of parameter photometric calibration 04\\
APPARE04 &$\Delta{\alpha_{{\rm ac},f}}$& mag\,mag$^{-1}$\,airmass$^{-1}$ & Error in parameter photometric calibration 04\\
APPN05   &                     &      & Name of parameter photometric calibration 05 (i.e., 'TimeTerm')\\
APPAR05  &$\alpha_{{\rm t},f}$         & mag\,day$^{-1}$ & Value of parameter photometric calibration 05\\
APPARE05 &$\Delta{\alpha_{{\rm t},f}}$ & mag\,day$^{-1}$ & Error in parameter photometric calibration 05\\
APPN06   &                     &      & Name of parameter photometric calibration 06 (i.e., 'TimeTerm2')\\
APPAR06  &$\alpha_{{\rm t2},f}$        & mag\,day$^{-2}$ & Value of parameter photometric calibration 06\\
APPARE06 &$\Delta{\alpha_{{\rm t2},f}}$& mag\,day$^{-2}$ & Error in parameter photometric calibration 06\\
APPN07   &                     &      & Name of parameter photometric calibration 07 (i.e., 'XTerm')\\
APPAR07  &$\alpha_{{\rm x1},f}$        & mag\,($X_{{\rm size}}$\,pix)$^{-1}$ & Value of parameter photometric calibration 07\\
APPARE07 &$\Delta{\alpha_{{\rm x1},f}}$& mag\,($X_{{\rm size}}$\,pix)$^{-1}$ & Error in parameter photometric calibration 07\\
APPN08   &                     &      & Name of parameter photometric calibration 08 (i.e., 'YTerm')\\
APPAR08  &$\alpha_{{\rm y1},f}$        & mag\,($Y_{{\rm size}}$\,pix)$^{-1}$ & Value of parameter photometric calibration 08\\
APPARE08 &$\Delta{\alpha_{{\rm y1},f}}$& mag\,($Y_{{\rm size}}$\,pix)$^{-1}$ & Error in parameter photometric calibration 08\\
APPN09   &                     &      & Name of parameter photometric calibration 09 (i.e., 'Y2Term')\\
APPAR09  &$\alpha_{{\rm y2},f}$        & mag\,($Y_{{\rm size}}$\,pix)$^{-2}$ & Value of parameter photometric calibration 09\\
APPARE09 &$\Delta{\alpha_{{\rm y2},f}}$& mag\,($Y_{{\rm size}}$\,pix)$^{-2}$ & Error in parameter photometric calibration 09\\
APPN10   &                     &      & Name of parameter photometric calibration 10 (i.e., 'Y3Term')\\
APPAR10  &$\alpha_{{\rm y3},f}$        & mag\,($Y_{{\rm size}}$\,pix)$^{-3}$ & Value of parameter photometric calibration 10\\
APPARE10 &$\Delta{\alpha_{{\rm y3},f}}$& mag\,($Y_{{\rm size}}$\,pix)$^{-3}$ & Error in parameter photometric calibration 10\\
APPN11   &                     &      & Name of parameter photometric calibration 11 (i.e., 'XYTerm')\\
APPAR11  &$\alpha_{{\rm xy},f}$        & mag\,($X_{{\rm size}} Y_{{\rm size}}$)$^{-1}$\,pix$^{-2}$ & Value of parameter photometric calibration 11\\
APPARE11 &$\Delta{\alpha_{{\rm xy},f}}$& mag\,($X_{{\rm size}} Y_{{\rm size}}$)$^{-1}$\,pix$^{-2}$ & Error in parameter photometric calibration 11 \\
\hline
{\it obsolete parameters} &                  &                                                      &  \\
PHTCALEX &                            &                                                      & Flag indicating if old photometric calibration routines were executed \\
PHTCALFL &                            &                                                      & Flag for image is photometric (0=N, 1=Y)\\
PCALRMSE &                            & mag                                                  & RMS of the residuals from (zeropoint, extinction) data fit.\\
IMAGEZPT &                            & mag                                                  & Image magnitude zeropoint\\
COLORTRM &                            & mag                                                  & Image ($g-r$) color term \\
ZPTSIGMA &                            & mag                                                  & Robust dispersion of SExtractor$-$SDSS magnitudes\\
IZPORIG  &                            &                                                      & Photometric-calibration catalog origin (e.g., SDSS) \\
ZPRULE   &                            &                                                      & Photometric-calibration method (e.g., DIRECT) \\
MAGZPT   &                            & mag                                                  & Magnitude zeropoint at airmass$=1$ \\
EXTINCT  &                            & mag\,airmass$^{-1}$                                   & Extinction 
\enddata
\tablecomments{FITS header keywords representing the nightly best-fit photometric calibration. In all cases the parameters are based on fitting Equation~\ref{ZPeqXY}.
Symbol represents the parameter designation used throughout this paper (e.g., Eqs~\ref{ZPeq_R}--\ref{ZPeqXY}).
The set of keywords below the horizontal line are obsolete parameters (see text).
}
\label{Tab:Header}
\end{deluxetable*}

The FITS headers contain additional parameters related to
photometric calibration.
However, these parameters are based on different routines
which are no longer supported.
For completeness, these parameters are listed in Table~\ref{Tab:Header}
under the obsolete parameters section.
These keywords are products of our first attempt
at photometric calibration of PTF images, which is less sophisticated
than the method presented here.
Both methods are still being run in the
pipeline and the results from both methods are still being written to the
FITS headers of the processed images.  A detailed description of
the first version of photometric calibration is beyond the scope of the
current paper, but will be available from online documentation.

\subsection{The illumination correction FITS image}
\label{ZPVM}

For each night, CCD and filter, we generate
two additional calibration products,
and package them in separate single-extension FITS-image files:
(i) the illumination correction image (or ZPVM image);
and (ii) an image of the standard deviation of the illumination correction.
Both images are created by linearly interpolating
the values of the means and standard deviations of the residuals
from Equation~\ref{ZPeq_R} or \ref{ZPeq_g} (see \S\ref{Process}).
The ZPVM image and associated standard deviation image are calculated
in units of magnitude.  The values of the ZPVM image are shifted
by a constant so that they are equal to zero
in the image center.
Setting the value at the image center to zero makes the
solutions based on Equations~\ref{ZPeq_R} or \ref{ZPeq_g} similar to
those based on Equation~\ref{ZPeqXY}.

The illumination correction images are not applied to
the individual processed PTF images.
One reason is that the illumination correction images
are reliable only for photometric nights.
The selection of photometric nights is discussed in \S\ref{Pars}.

\subsection{Use of data products}
\label{Use}

Each processed PTF image is accompanied by an associated
FITS binary table containing a catalog of
objects extracted from the processed image by SExtractor.
The catalog
source magnitudes are instrumental magnitudes, and users of this
product can convert instrumental magnitudes into calibrated
magnitudes using the provided PTF photometric parameters and the
following procedure.
In order to convert the MAG\_AUTO\footnote{For description of the MAG\_AUTO magnitude see the SExtractor manual: http://www.astromatic.net/software/sextractor.}
instrumental magnitudes to
calibrated magnitudes, one needs to apply Equation~\ref{ZPeq_R} or \ref{ZPeq_g}
and add the value of the ZPVM image at the location of the object.
Alternatively, it is possible to directly use Equation~\ref{ZPeqXY}.

Applying Equation~\ref{ZPeq_R}, \ref{ZPeq_g} or \ref{ZPeqXY} requires
the colors of individual sources in the SDSS system.
If the color of a source is known, from PTF
observations or other surveys (e.g., SDSS),
then it is possible to use these equations
to get the magnitude of that source in the SDSS system.
However, not all PTF fields were observed in both
the $g$ and $R$ filters or are within the SDSS footprint.
In these cases, 
setting the color terms $\alpha_{{\rm c},f}$ and $\alpha_{{\rm ac},f}$ to zero
will return
the object's magnitude in the {\it non-color-corrected PTF magnitude system} (i.e., the natural PTF magnitude system).
We denote magnitudes in this system by $f_{{\rm PTF}}$, where $f$ is the filter moniker.
PTF magnitudes which are color-corrected (i.e., the source color is
known and applied to Equations~\ref{ZPeq_R}, \ref{ZPeq_g} or \ref{ZPeqXY})
are denoted by $f_{{\rm PTF/SDSS}}$.
We note that applying Equations~\ref{ZPeq_R}, \ref{ZPeq_g} or \ref{ZPeqXY}
is done by isolating $f_{{\rm SDSS}}$ and replacing it with
$f_{{\rm PTF/SDSS}}$ in cases where the color is known,
or $f_{{\rm PTF}}$ if the colors are set to zero.
We note that the calibrated $f_{{\rm PTF}}$ magnitudes of all sources
detected in PTF images are available from the PTF database.

As in the case of any other magnitude system,
in order to convert the PTF non-color corrected magnitudes
to other systems, one needs to know the color index of the source.
In some cases, it is possible to use some prior knowledge
about the mean color index of a class of objects
(e.g., RR Lyr stars; asteroids),
in order to convert their PTF magnitudes (in a single band) into SDSS magnitudes.
For example, 68\% of main-belt asteroids have $r-i$ color index
in a narrow range of about 0.2\,mag (e.g., Ivezi\'{c} et al. 2001).
Therefore, given the PTF $R$-band color coefficients,
$\alpha_{{\rm c},R}\approx0.2$,
the scatter in the mean color correction for asteroids
is about 0.04\,mag.
Neglecting the airmass-color terms (which are typically small; see Table~\ref{Tab:Terms}),
the relations between the PTF and the SDSS magnitude systems,
for objects with stellar-like spectra, are given by:
\begin{equation}
r_{{\rm SDSS}} - R_{{\rm PTF}}  \cong \alpha_{{\rm c},R} (r_{{\rm SDSS}} - i_{{\rm SDSS}}),
\label{rPTF_SDSS}
\end{equation}
and
\begin{equation}
g_{{\rm SDSS}} - g_{{\rm PTF}}  \cong \alpha_{{\rm c},g} (g_{{\rm SDSS}} - r_{{\rm SDSS}}),
\label{gPTF_SDSS}
\end{equation}
where the values of the color terms are listed in Table~\ref{Tab:Terms}.

In order to apply the illumination correction
to individual images, we need to multiply,
pixel by pixel, the processed image
by $10^{-0.4\times {\rm ZPVM}}$,
where ZPVM is the illumination correction image.
Alternatively, one can use the less accurate polynomial representation
of the ZPVM image.

\section{Performance}
\label{Perf}


The accuracy of our calibration process, relative to SDSS,
is discussed in \S\ref{Acc}, while in \S\ref{Rep},
we discuss the repeatability of our photometric calibration.

\subsection{Accuracy}
\label{Acc}

As described in \S\ref{Use},
the PTF calibrated magnitudes can be given
in two systems.
A color-corrected magnitude system,
denoted by $R_{{\rm PTF/SDSS}}$ or $g_{{\rm PTF/SDSS}}$,
in which the known color of the object is
applied to Equations~\ref{ZPeq_R}--\ref{ZPeq_g} or \ref{ZPeqXY}.
By construction, this magnitude system is similar to the SDSS system.
The other possibility is a ``natural system''
in which all the color terms involving
$\alpha_{{\rm c},f}$ and $\alpha_{{\rm ac},f}$
are set to zero (i.e., the color indices of the objects are ignored).
These magnitudes are denoted by
$R_{{\rm PTF}}$ or $g_{{\rm PTF}}$.

\begin{figure}
\centerline{\includegraphics[width=8.5cm]{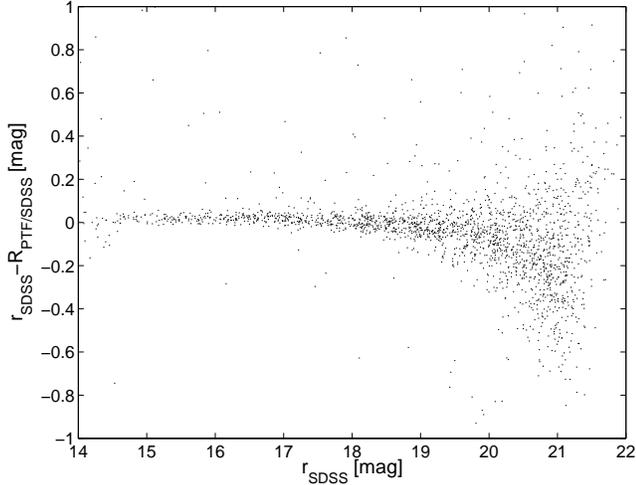}}
\caption{$r_{{\rm SDSS}}-R_{{\rm PTF/SDSS}}$ as a function of
SDSS $r$-band magnitude ($r_{{\rm SDSS}}$) for point sources
in a single PTF CCD image.
The image was taken on 2010 March 18.3696 with CCDID$=0$
were the PTF camera was centered on PTF field 2961.
\label{fig:ColorCorr}}
\end{figure}

\begin{figure}
centerline{\includegraphics[width=8.5cm]{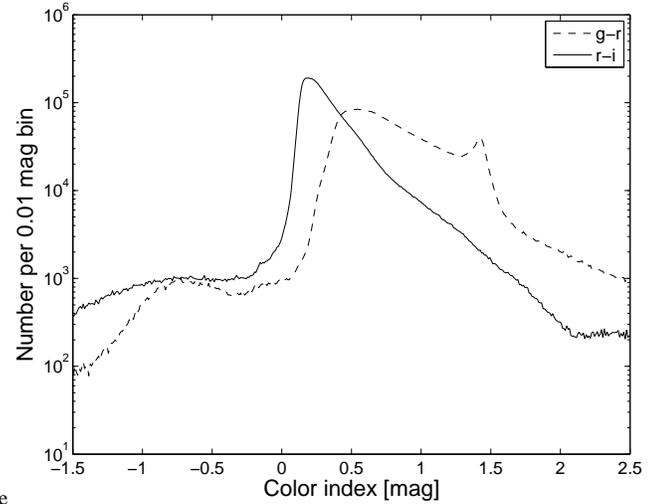}}
\caption{The $g-r$ and $r-i$ color distributions of all SDSS-DR7 point sources
with $r$-band magnitude between 16 and 17.
\label{fig:SDSS_color_dist_r16_17_stars}}
\end{figure}

To demonstrate the current accuracy of the PTF
photometric calibration, we show in Figure~\ref{fig:ColorCorr}
the differences between SDSS magnitudes and
color-corrected PTF $R$-band magnitudes ($R_{{\rm PTF/SDSS}}$),
as a function of SDSS magnitude.
%
\begin{figure}
\centerline{\includegraphics[width=8.5cm]{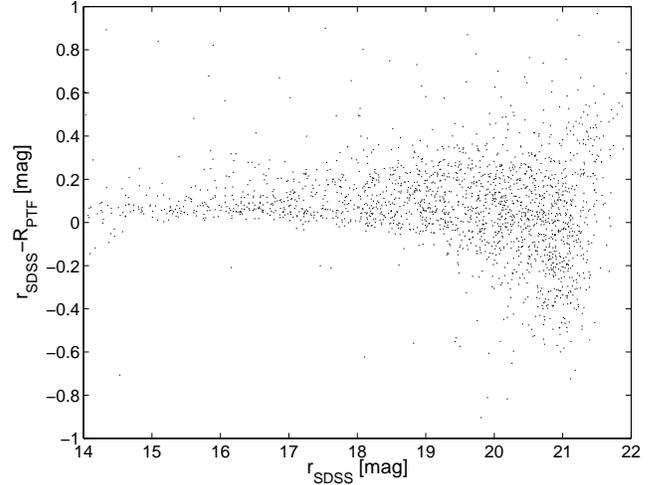}}
\caption{Same as Figure~\ref{fig:ColorCorr} but
without applying the color terms to the PTF magnitudes
(i.e., assuming all the stars have $r-i=0$\,mag).
\label{fig:ColorUnCorr}}
\end{figure}
In this particular example, the StD
in the $r_{SDSS}-R_{PTF/SDSS}$ for bright stars
is smaller than $\approx25$\,mmag.
Figure~\ref{fig:ColorUnCorr} shows the same, but for PTF magnitudes
not corrected for the color term ($R_{{\rm PTF}}$).
The scatter in the uncorrected color plot
is larger, as expected.
The larger scatter in Fig.~\ref{fig:ColorUnCorr} is due to the fact
that different stars in the field of view have
different colors.
The 95th percentile range of $r-i$ color
of SDSS stars in the $r$-band magnitude range of 16 to 17
is between $-0.2$ and $1.2$\,mag (see Fig.~\ref{fig:SDSS_color_dist_r16_17_stars}).
Given that the PTF mean $r-i$ color term is about $0.2$\,mag,
this corresponds to a color correction 
between $-0.04$ and $0.24$\,mag.

We note that in Figures~\ref{fig:ColorCorr}-\ref{fig:ColorUnCorr},
the magnitude difference is systematically tending toward larger negative
values, starting around magnitude 19.5.
This is due to the use of MAG\_AUTO.
Indeed, using aperture magnitude
(i.e., MAG\_APER), we do~not see
any evidence for this bias.
This is because MAG\_AUTO adjusts the
aperture used to extract the source magnitude for each object.
Therefore, if MAG\_AUTO is used for sources
near the survey detection limit it is recommended
to estimate the aperture correction
for these sources
in order to obtain their unbiased magnitudes.
We note that the MAG\_AUTO is different from
aperture or PSF magnitudes.
In order to convert between MAG\_AUTO and other kinds of magnitude
estimators, one needs to find the aperture correction between
these magnitudes.
Both MAG\_AUTO and aperture magnitudes in predefined five apertures
are available in the FITS binary source catalog associated with each image.
Future versions of the PTF photometric pipeline will use
aperture or PSF magnitudes.

\subsection{Repeatability}
\label{Rep}

\begin{figure}
\centerline{\includegraphics[width=8.5cm]{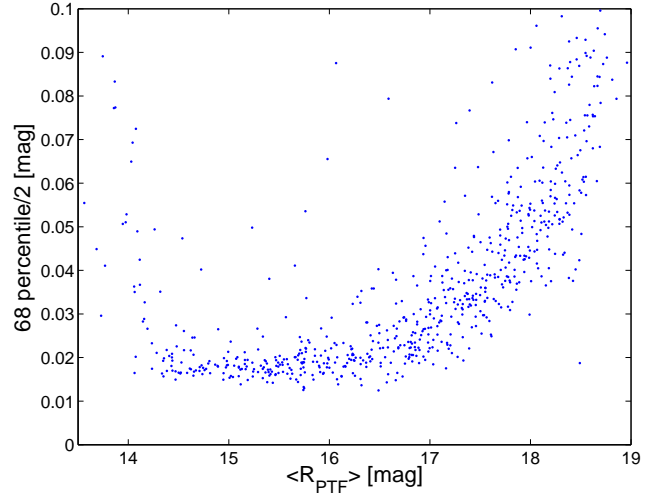}}
\caption{$R$-band magnitude vs. scatter for stars in PTF field 100037 CCDID$=0$.
The plot shows the 68th percentile range divided by 2
as a function of the PTF mean magnitude (not color corrected).
The 68th percentile range, for each source, is calculated over 27 images
of this field taken between March and October 2010
which have photometric calibration bright-star RMS value
(see definition for parameter APBSRMS in Table~\ref{Tab:Header})
of less than 0.04\,mag.
We note that the increase in scatter at the bright end is due to
saturated stars.
\label{fig:F100037_0_CalibMag_RMS}}
\end{figure}

The stability over time of the PTF $R$-band photometric solutions
is demonstrated in Figure~\ref{fig:F100037_0_CalibMag_RMS}.
This figure shows the scatter in the calibrated magnitude of
each star, over multiple epochs, as a function
of magnitude.
The scatter is calculated using the 68th percentile range divided by two
(i.e., equivalent to StD in case of a Gaussian distribution).
This is based on 27 images of
PTF field\footnote{PTF field (denoted by PTFFIELD) is a
an internal index associated with a unique field position on the celestial sphere, where PTF fields are mostly non-overlapping.}
100037 CCDID\footnote{CCD number is denoted by CCDID and runs from 0 to 11.}$=0$ 
taken between March and October 2010,
which have photometric calibration bright-star root-mean squares (RMS) value
(parameter APBSRMS in Table~\ref{Tab:Header})
of less than 0.04\,mag.

We choose to use the 68th percentile range
instead of StD since it is more robust to outliers.
For example, if a night was ``photometric'' for
90\% of the time (e.g., clouds entered only toward the end of the night),
then our pipeline might claim that the night was photometric, but the calibration
of some of the data will be poor.
Therefore, in order to get the calibrated source magnitudes,
it is important to average the data taken over several
photometric nights.
We note that we are using this approach for the compilation
of the PTF photometric catalog (Ofek et al., in prep.).
In turn, this catalog will be used in later phases
to calibrate the entire PTF data.

\begin{figure}
\centerline{\includegraphics[width=8.5cm]{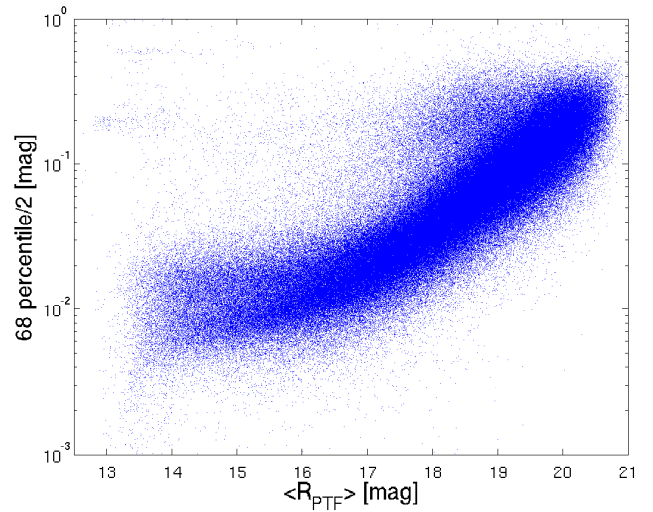}}
\caption{Like Figure~\ref{fig:F100037_0_CalibMag_RMS},
but for stars in 100 representative PTF fields/CCDs
which were observed at least on three photometric nights.
\label{fig:Mag_Scatter_100}}
\end{figure}

Figure~\ref{fig:F100037_0_CalibMag_RMS} suggests that,
at the bright end, the repeatability
of PTF calibrated magnitudes is good to a level of 0.02\,mag.
Figure~\ref{fig:Mag_Scatter_100} shows the same, but
for stars in 100 representative PTF fields and all CCDs.
The typical repeatability of PTF magnitudes is field-dependent.
The median repeatability over all these 100 fields, for all CCDs,
for stars between 15 to 15.5\,mag, is
about $11$\,mmag and the 95-percentile range is
between 2\,mmag to 44\,mmag.
We note that PTF $g$-band observations achieve similar
repeatability.

\section{Photometric parameter statistics}
\label{Pars}

\begin{figure}
\centerline{\includegraphics[width=8.5cm]{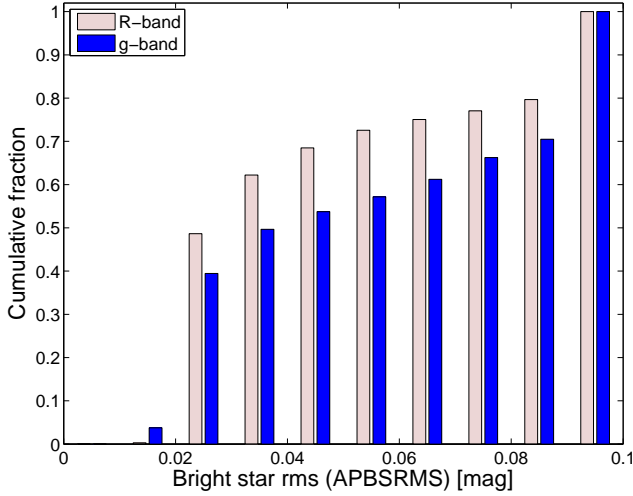}}
\caption{Cumulative histogram of the bright-star RMS parameter
(i.e., observations that have bright-star RMS scatter below a particular value)
for all PTF $R$-band (red) and $g$-band (blue) images.
\label{fig:Hist_APBSRMS}}
\end{figure}

\begin{figure}
\centerline{\includegraphics[width=8.5cm]{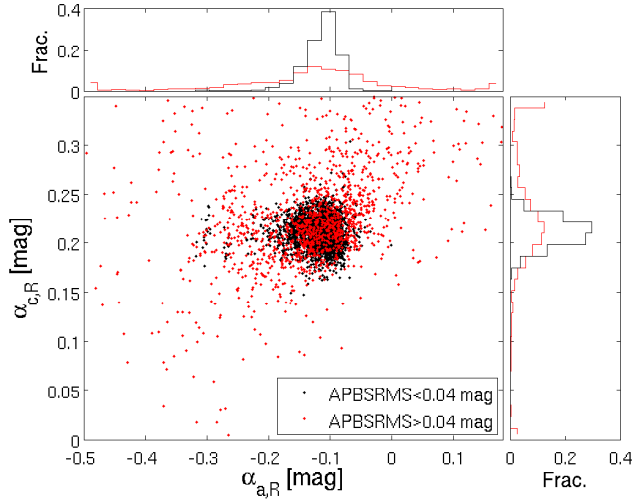}}
\caption{Airmass term vs. color term of all the PTF $R$-band data in all CCDs.
Black points represent images taken during nights in which the bright-star
RMS value was smaller than 0.04\,mag, while the rest of the data is marked in red points.
The histograms of airmass and color terms are shown on the top and right of the figure,
respectively. The histograms are normalized by the sum of the number of data points.
\label{fig:alphaCr_alphaAr}}
\end{figure}

\begin{figure}
\centerline{\includegraphics[width=8.5cm]{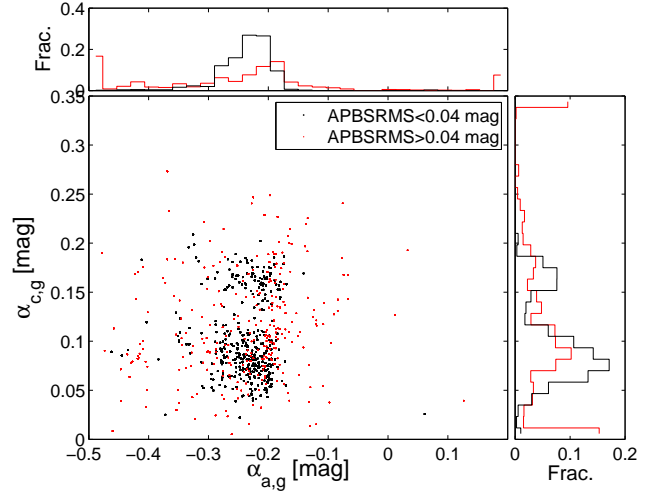}}
\caption{The same as Figure~\ref{fig:alphaCr_alphaAr} but for $g$-band.
The bimodality is due to different blue-side spectral response of the various CCDs
in the PTF camera (see Table~\ref{Tab:Terms}).
We note that most of the PTF data were taken in $R$-band.
\label{fig:alphaCg_alphaAg}}
\end{figure}

The term ``photometric night'' is not well defined
and depends on the required accuracy.
Figure~\ref{fig:Hist_APBSRMS} shows the cumulative histogram
of the RMS of the best-fit residuals of bright stars
(parameter APBSRMS in Table~\ref{Tab:Header})
in all the images taken by PTF.
This figure suggests that about 62\% (50\%) of the $R$-band ($g$-band)
images taken by PTF
have photometric calibration with bright-star RMS values smaller
than about 0.04\,magnitude.
Figure~\ref{fig:alphaCr_alphaAr} shows the $R$-band airmass term ($\alpha_{{\rm a},R}$)
vs. the color term ($\alpha_{{\rm c},R}$), while
Figure~\ref{fig:alphaCg_alphaAg} shows the same for $g$-band.
These figures suggest that {\it some} nights that have
small bright-star RMS values (i.e., smaller than 0.04\,mag)
have anomalous airmass or/and color coefficients.
Therefore, 
it is better to select data taken in nights
for which the airmass and color terms are close
to their average values (e.g., see Table~\ref{Tab:Terms}).
Here we define photometric nights as
nights with bright-star RMS value smaller than 0.04\,mag
and $\alpha_{{\rm c},f}$ 
within the range 
$\langle\alpha_{{\rm c},f}\rangle \pm 3 \Delta[\alpha_{{\rm c},f}]$
and  $\alpha_{{\rm a},R}$
within the range 
$\langle\alpha_{{\rm a},f}\rangle \pm 3 \Delta[\alpha_{{\rm a},f}]$.
Here, 
$\langle\alpha_{{\rm c},f}\rangle$,
$\Delta[\alpha_{{\rm c},f}]$,
$\langle\alpha_{{\rm a},f}\rangle$,
and 
$\Delta[\alpha_{{\rm a},f}]$
are the median and 1-$\sigma$ width
of the color and extinction coefficients
for each CCD as listed in Table~\ref{Tab:Terms}.
We find that 59\% (47\%) of the $R$-band ($g$-band)
PTF data were taken in such photometric nights.

\begin{figure}
\centerline{\includegraphics[width=8.5cm]{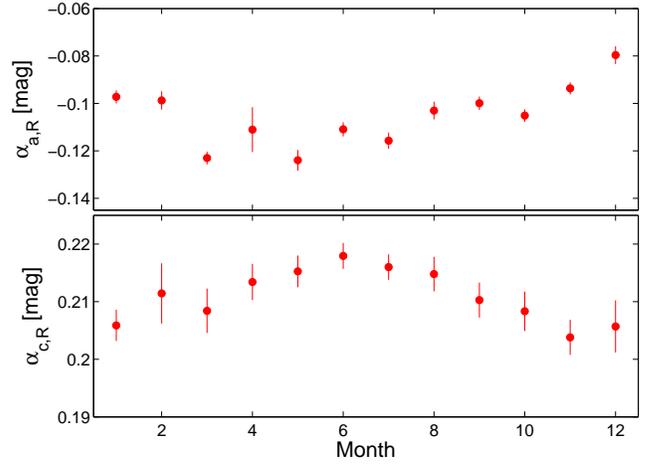}}
\caption{Seasonal variations in $\alpha_{{\rm a},R}$ and $\alpha_{{\rm c},R}$
as a function of month. This plot is based on data taken between
May 2010 and May 2011.
\label{fig:Alpha_seasonal}}
\end{figure}

The atmospheric extinction coefficients 
depend on the atmospheric conditions.
Therefore, they may vary with season.
Figure~\ref{fig:Alpha_seasonal} shows the median $R$-band airmass coefficient ($\alpha_{{\rm a},R}$)
and the color term ($\alpha_{{\rm c},R}$)
as a function of month in the year.
The error bars are given by the StD within each month
divided by the square root of the number of photometric
nights in each month.
This figure indicates that there are small, but significant,
seasonal variations in both the $R$-band extinction and
color terms.

Table~\ref{Tab:Terms} presents the median and
68th percentile range divided by 2,
of the color-term, extinction-coefficient
and airmass-color terms for each CCD and filter.
These are calculated based on data from all nights 
in which the RMS of the best-fit residuals of bright stars is smaller than 0.04\,mag and the
limiting magnitude\footnote{The limiting magnitude is estimated
based on the calibrated photometric parameters, the sky background
and assuming a 3-$\sigma$ detection limit. This property is stored in the image headers, as well as in the PTF database.}
is fainter than 20.5.
We note that the $R$-band color terms
are roughly the same for all CCDs, while
the $g$-band color terms have a wider range.
\begin{deluxetable*}{lrlrlrlrr}
\tablecolumns{9}
\tabletypesize{\scriptsize}
\tablewidth{0pt}
\tablecaption{Median photometric-calibration parameters and associated parameter dispersion}
\tablehead{
\colhead{CCDID} &
\colhead{$\langle\alpha_{{\rm c},f}\rangle$} &
\colhead{$\Delta[\alpha_{{\rm c},f}]$} &
\colhead{$\langle\alpha_{{\rm a},f}\rangle$} &
\colhead{$\Delta[\alpha_{{\rm a},f}]$} &
\colhead{$\langle\alpha_{{\rm ac},f}\rangle$} &
\colhead{$\Delta[\alpha_{{\rm ac},f}]$} &
\colhead{$\langle T_{1,f}\rangle$} &
\colhead{$\langle T_{2,f}\rangle$} 
}
\startdata
$f=g$-band filter & &    &          &         &          &         &           & \\
00   &$  0.152$&$  0.017$&$  -0.226$&$  0.038$&$   0.016$&$  0.012$&$    0.704$&$ -0.109$\\
01   &$  0.075$&$  0.010$&$  -0.227$&$  0.035$&$   0.020$&$  0.008$&$    0.832$&$ -0.097$\\
02   &$  0.100$&$  0.016$&$  -0.234$&$  0.037$&$   0.014$&$  0.011$&$    0.790$&$ -0.101$\\
04   &$  0.095$&$  0.013$&$  -0.238$&$  0.035$&$   0.015$&$  0.009$&$    0.798$&$ -0.100$\\
05   &$  0.169$&$  0.009$&$  -0.222$&$  0.032$&$   0.008$&$  0.007$&$    0.676$&$ -0.111$\\
06   &$  0.075$&$  0.022$&$  -0.233$&$  0.038$&$   0.022$&$  0.012$&$    0.832$&$ -0.097$\\
07   &$  0.068$&$  0.017$&$  -0.237$&$  0.038$&$   0.018$&$  0.013$&$    0.844$&$ -0.096$\\
08   &$  0.075$&$  0.013$&$  -0.243$&$  0.041$&$   0.016$&$  0.007$&$    0.832$&$ -0.097$\\
09   &$  0.085$&$  0.012$&$  -0.231$&$  0.036$&$   0.019$&$  0.008$&$    0.816$&$ -0.098$\\
10   &$  0.075$&$  0.020$&$  -0.236$&$  0.037$&$   0.017$&$  0.013$&$    0.831$&$ -0.097$\\
11   &$  0.161$&$  0.015$&$  -0.225$&$  0.041$&$   0.011$&$  0.011$&$    0.689$&$ -0.110$\\
All  &$  0.090$&$  0.047$&$  -0.234$&$  0.037$&$   0.016$&$  0.011$&$    0.806$&$ -0.099$\\
\hline
$f=R$-band filter & &    &          &         &          &         &           & \\
00   &$  0.217$&$  0.009$&$  -0.114$&$  0.023$&$   0.006$&$  0.006$&$    0.080$&$  0.125$\\
01   &$  0.206$&$  0.010$&$  -0.109$&$  0.023$&$   0.008$&$  0.006$&$    0.093$&$  0.122$\\
02   &$  0.202$&$  0.008$&$  -0.112$&$  0.025$&$   0.007$&$  0.006$&$    0.096$&$  0.121$\\
04   &$  0.219$&$  0.010$&$  -0.119$&$  0.027$&$   0.003$&$  0.006$&$    0.079$&$  0.125$\\
05   &$  0.227$&$  0.009$&$  -0.117$&$  0.027$&$   0.002$&$  0.007$&$    0.070$&$  0.127$\\
06   &$  0.225$&$  0.008$&$  -0.106$&$  0.026$&$   0.005$&$  0.006$&$    0.072$&$  0.127$\\
07   &$  0.205$&$  0.010$&$  -0.105$&$  0.023$&$   0.008$&$  0.007$&$    0.093$&$  0.122$\\
08   &$  0.205$&$  0.008$&$  -0.112$&$  0.021$&$   0.005$&$  0.005$&$    0.093$&$  0.122$\\
09   &$  0.201$&$  0.012$&$  -0.113$&$  0.026$&$   0.004$&$  0.008$&$    0.097$&$  0.121$\\
10   &$  0.218$&$  0.014$&$  -0.113$&$  0.025$&$   0.004$&$  0.010$&$    0.080$&$  0.125$\\
11   &$  0.218$&$  0.010$&$  -0.114$&$  0.025$&$   0.003$&$  0.007$&$    0.080$&$  0.125$\\
All  &$  0.212$&$  0.014$&$  -0.112$&$  0.025$&$   0.005$&$  0.007$&$    0.086$&$  0.124$
\enddata
\tablecomments{The first column indicates the CCDID, where ``All'' represents
all the CCDs. Parameters $T_{1}$ and $T_{2}$ for band $f$ are defined  in Equations~\ref{RptfRc}--\ref{gptfV}. $\Delta$ represents the 68th percentile range divided by 2,
a robust measure of scatter in a parameter.}
\label{Tab:Terms}
\end{deluxetable*}

\section{Color transformations}
\label{Color}

Approximate relations between the non-color-corrected PTF magnitude system
and the SDSS magnitude system are given in
Equations~\ref{rPTF_SDSS}--\ref{gPTF_SDSS}.
The exact relations between the two systems depend on the details
of the object's spectral energy distribution and the
atmospheric conditions at the time of the observations.

Using Equation~\ref{rPTF_SDSS} and the color transformations
between the Johnson-Cousins ($UBVR_{{\rm c}}I_{{\rm c}}$) system and the SDSS system\footnote{http://www.sdss.org/dr6/algorithms/sdssUBVRITransform.html.},
we find an approximate
transformation between PTF $R$-band magnitude and the Johnson-Cousins system:
\begin{eqnarray}
R_{{\rm PTF}} &\cong & R_{{\rm c}}+(R_{{\rm c}}-I_{{\rm c}})(0.3088-1.0517\alpha_{{\rm c},R}) \cr
            &      & + 0.0704 + 0.2504\alpha_{{\rm c},R} \cr
            &\equiv& T_{{\rm 1},R}(R_{{\rm c}}-I_{{\rm c}}) + T_{{\rm 2},R},
\label{RptfRc}
\end{eqnarray}
and similarly using Equation~\ref{gPTF_SDSS} we get for PTF $g$-band:
\begin{eqnarray}
g_{{\rm PTF}} &\cong & V+(V-R_{{\rm c}})(0.9556-1.6521\alpha_{{\rm c},g}) \cr
            &      & - 0.0853 - 0.1541\alpha_{{\rm c},g} \cr
            &\equiv& T_{{\rm 1},g}(V-R_{{\rm c}}) + T_{{\rm 2},g}.
\label{gptfV}
\end{eqnarray}
Here $T_{{\rm 1},f}$ and $T_{{\rm 2},f}$ are the
color-transformation coefficients
for filter $f$. These parameters are listed in Table~\ref{Tab:Terms}.

Assuming\footnote{http://www.sdss.org/dr6/algorithms/sdssUBVRITransform.html.}
$V=+0.03$ and $V-R_{{\rm c}}=R_{{\rm c}}-I_{{\rm c}}=0$,
we find for the A0V star Vega:
\begin{equation}
R_{{\rm PTF}}^{{\rm Vega}} \cong 0.17 + 0.23 \alpha_{{\rm c},R},
\label{RptfVega}
\end{equation}
and
\begin{equation}
g_{{\rm PTF}}^{{\rm Vega}} \cong -0.08 + 0.25 \alpha_{{\rm c},g}.
\label{RptfVega}
\end{equation}

For the absolute magnitude of the Sun,
assuming $V=+4.82$, $V-R_{{\rm c}}=+0.36$, $R_{{\rm c}}-I_{{\rm c}}=+0.32$
and using SDSS color transformations,
we find:
\begin{equation}
R_{{\rm PTF}}^{{\rm Sun}} \cong 4.68 - 0.11 \alpha_{{\rm c},R},
\label{RptfVega}
\end{equation}
and
\begin{equation}
g_{{\rm PTF}}^{{\rm Sun}} \cong 5.12 -0.44 \alpha_{{\rm c},g}.
\label{RptfVega}
\end{equation}

\section{Conclusion}
\label{Sum}

A wide variety of scientific applications
require photometric calibration of astronomical
images.
In this paper, we provide a description
of the current PTF photometric calibration process and products.
We show that roughly half of the PTF data were taken
under conditions that allow calibration of the images
to an accuracy of 0.02--0.04\,mag relative to SDSS.
Most importantly, the rest of the data, taken under non-photometric conditions,
can be calibrated using data taken in photometric conditions.
At the bright end ($\approx15$\,mag), the PTF photometric calibration is stable
to a level of $\approx0.02$\,mag.
Future improvements to the PTF photometric calibration
process that are important for the PTF
type-Ia and IIp supernovae cosmology projects
will be described elsewhere.

The PTF photometric calibration allows us to measure
the calibrated magnitude of transients outside the SDSS
footprint.
We have used the PTF dataset to compile a photometric
catalog.
The first version of this catalog is described in Ofek et al. (in prep.).
We note that our tests show that
PTF data can deliver 
relative photometry, in single images, with precision as good as 3\,mmag
(e.g., van Eyken et al. 2011; Ag\"{u}eros et al. 2011).
This capability is being built into the PTF relative photometry pipeline
and will be described in Levitan et al. (in prep.).

\acknowledgments

We thank Andrew Pickles and an anonymous referee
for useful comments on the manuscript.
This paper is based on observations obtained with the
Samuel Oschin Telescope as part of the Palomar Transient Factory
project, a scientific collaboration between the
California Institute of Technology,
Columbia University,
Las Cumbres Observatory,
the Lawrence Berkeley National Laboratory,
the National Energy Research Scientific Computing Center,
the University of Oxford, and the Weizmann Institute of Science.
EOO is supported by an Einstein fellowship and NASA grants.
SRK and his group are partially supported by the
NSF grant AST-0507734.
SBC acknowledges generous financial assistance from
Gary \& Cynthia Bengier, the Richard \& Rhoda Goldman Fund,
NASA/{\it Swift} grants NNX10AI21G and GO-7100028,
the TABASGO Foundation, and NSF grant AST-0908886.

\end{document}